\documentclass[onecolumn,nofootinbib]{revtex4}
\usepackage{graphicx}
\usepackage{amsmath}
\usepackage{amssymb}
\usepackage{float}
\usepackage[all]{xy}
\usepackage{amsfonts}
\usepackage{dcolumn}
\usepackage{bm}
\usepackage{xcolor}
\usepackage{caption}
\usepackage{hyperref}
\usepackage{multirow}
\usepackage[utf8x]{inputenc}
\usepackage{epstopdf}
\usepackage{capt-of}
\usepackage{dcolumn}   
\usepackage{bm}
\usepackage{dsfont}
\usepackage{relsize}
\usepackage{mathtools}
\usepackage{xparse}

\begin{document}

\title{Constraining $f(T,B)$ Teleparallel Gravity From Energy Conditions}

\author{Snehasish Bhattacharjee}
\email{snehasish.bhattacharjee.666@gmail.com}

\affiliation{Department of Astronomy, Osmania University, Hyderabad, 500007, India}
\date{\today}

\begin{abstract}
\textbf{Abstract:} $f(T,B)$ teleparallel gravity is a recently proposed straightforward generalization of the popular $f(T)$ teleparallel gravity by the incorporation of a boundary term $B=\frac{2}{e}\partial_{i}(e T ^{i}) = \bigtriangledown_{i}T^{i}$ where $T$ denote the torsion scalar \cite{ftb13}. In this work, I investigate the viability of some well motivated $f(T,B)$ teleparallel gravity models of the forms $f=\alpha  B^n+\beta  T^m$, $f=\alpha  B^n T^m$ and $f=\alpha  \log (B)+\beta  T$ where $\alpha, \beta, n$ and $m$ are free parameters from the inequalities imposed the the weak energy condition. I use the recent estimates of Hubble, deceleration, jerk and snap parameters in finding corners in parameter spaces for the cosmological models for which the energy density remain positive and the weak energy condition ( i.e, $\rho+p \geq 0$, where $p$ and $\rho$ represent respectively the cosmological pressure and energy density) attains a minute positive value, as this implies the EoS parameter $\omega = p/\rho \simeq-1$ and therefore consistent with an accelerating universe. 

\end{abstract}

\maketitle
\section{Introduction}
Cosmological observations indicate the universe is accelerating \cite{observations}. Understanding what fuels this late-time acceleration is one of the most important problems in theoretical physics. Additionally, the formation of large scale structures requires the presence of supplementary non-dissipative matter in high abundance. These cosmological entities together account for about 95\% of the energy budget of the universe but hitherto has no observational evidence to support their holy presence. Several candidates to expound these phenomena have been reported but none as popular as dark matter and dark energy. \\
The physics governing dark energy has very little theoretical motivation since it possesses a negative EoS parameter and thus create a repulsive gravitational force which is in total contrast with the usual attractive nature of gravity. Interestingly, there are two ways to circumvent this problem: Firstly, one can introduce extra scalar fields which possess a negative EoS parameter adhered to general relativity and secondly, modify general relativity at high energy scales which results in some kind of dark energy effects purely geometrical in nature (see \cite{ref3,ref4,ref5} for some recent interesting results in modified gravity theories). \\
Amongst numerous modified gravity theories, teleparallel gravity theory has gained significant prominence in addressing the late-time acceleration \cite{ftb17,mandal,cap2}. Recently, teleparallel gravity has been applied and yielded interesting results in gravitational baryogenesis \cite{fqt,baryo/2020}. Cosmography in $f(T)$ gravity was studied in \cite{cap1}. In \cite{ftb13,no}, the simple $f(T)$ teleparallel gravity was generalized through the introduction of a new Lagrangian $f(T,B)$, where $B$ is a boundary term related to the divergence of torsion scalar $T$. Thermodynamical studies in $f(T,B)$ gravity were reported in \cite{ftb}. Interestingly for $f(-T+B)$, the field equations correspond to that of $f(R)$ gravity \cite{cap3}. In this work, I am interested in constraining some widely used $f(T,B) $ gravity models in a flat FRW spacetime with the energy conditions. Particularly, I first presume the cosmic fluid obeys the equation of state of the form $p=\omega \rho$ where $p$, $\rho$ and $\omega$ represent respectively the cosmological pressure, energy density and EoS parameter. I then find corners in parameter spaces for which the energy density remains positive and the weak energy condition is slightly greater than zero as this implies $\omega \simeq -1$ and therefore consistent with latest Planck measurements \cite{planck}. \\Energy conditions are crucial for understanding the physics of singularities and classical black holes thermodynamics (see \cite{hawking} for more details). One can construct the energy conditions from the Raychaudhuri equation for an expanding universe, where the attractive nature of gravity indicate $R_{ij}\kappa^{i}\kappa^{j}\geq 0$, where $R_{ij}$ represents the Ricci tensor and $\kappa^{i}$ any null vector. In general relativity, one obtains $T_{ij}\kappa^{i}\kappa^{j}\geq 0$, where $T_{ij}$ represents the stress energy momentum tensor. This inequality is commonly known as the null energy condition. The weak energy condition postulates that the local energy density is always positive and therefore $T_{ij}U^{i}U^{j}\geq0$, for all timelike vectors $U^{i}$. In this paper, I focus on the case where the cosmic fluid is perfect and therefore the weak energy condition can be re-written in simplified form as $\rho\geq0$ and $\rho+p\geq0$. Energy conditions have been extensively used to constrain $f(R)$ gravity \cite{ecfr}, $f(T)$ gravity \cite{ecft}, $f(G)$ gravity \cite{ecfg}, $f(R,G)$ gravity \cite{ecfrg} and $f(R,T)$ gravity \cite{ecfrt}. However, as far as I know, no studies aimed at constraining $f(T,B)$ gravity has been made so far. Since this is a newly constructed modified gravity theory, the constraints imposed by energy conditions would be very useful in understanding the efficiency and applicability of this modified gravity in cosmology. \\The manuscript is organized as follows: In Section \ref{II}, I summarize the $f(T,B)$ teleparallel gravity. In Section \ref{III} I describe the and obtain expressions of the energy conditions from the $f(T,B)$ gravity Friedmann equations. In Section \ref{IV} I present some well motivated $f(T,B)$ gravity models and put constraints on the parameter spaces for which the weak energy condition attains a small positive value and in Section \ref{V} I present the conclusions and final remarks.

\section{ $f(T,B)$ Gravity}\label{II}
In this section I shall briefly discuss teleparallel $f(T)$ gravity and its extension to $f(T,B)$ gravity. In this theory, the tetrads $e^{a}_{i}$ are the dynamical variables which form an orthonormal basis for the tangent space at each point $x^{i}$ of the spacetime manifold \cite{no}. Therefore, $e^{m}_{i}$ along with their inverses $U^{i}_{n}$ follow:
\begin{equation}
e^{m}_{i}E_{n}^{i}=\delta^{m}_{n}, \hspace{0.25in} e^{n}_{j}E_{n}^{j}=\delta^{j}_{i}.
\end{equation}
From these relations, the metric $g_{ij}$ can be written as 
\begin{equation}
g_{ij}=e^{a}_{i}e^{b}_{j}\eta_{ab},
\end{equation}
where $\eta_{ab}$ represents the Minkowski metric. In teleparallel gravity, the geometry is coupled with a torsion and is globally curvature less. To achieve this, the torsion tensor is defined as
\begin{equation}
T^{a}_{ij} = \partial_{i}e^{a}_{j}-\partial_{j}e^{a}_{j}.
\end{equation}
Furthermore, the contorsion tensor reads
\begin{equation}
T^{\epsilon}{}_{ij}-T_{ij}{}^{\epsilon}+T_{i}{}^\epsilon{}_{j}=2 F_{i}{}{\epsilon}{}_{j}
\end{equation}
and also the following tensor 
\begin{equation}
2C_{\epsilon}{}^{ij} = F_{\epsilon}{}^{ij}-\partial^{i}_{\epsilon}T^{i}+\partial^{j}_{\epsilon}T^{j}.
\end{equation}
The joint term $C_{\epsilon}{}^{ij}T^{\epsilon}{}_{ij} $ is called the torsion scalar $T$. Remarkably the Ricci scalar $R$ can be defined as 
\begin{equation}
R =-T + \frac{2}{e}\partial_{i}(e T ^{i})=-T+B.
\end{equation}
Where $B$ is nothing but a boundary term defined as $B=\frac{2}{e}\partial_{i}(e T ^{i}) = \bigtriangledown_{i}T^{i}$. Thus teleparallel gravity coupled with $B$ reproduces the exact field equations as in general relativity. \\
With that reasoning, the authors in \cite{ftb13} introduced the $f(T,B)$ teleparallel gravity in which the action is defined as 
\begin{equation}\label{1}
\mathcal{S} = \frac{1}{\kappa}\int d^{4}x e f(T,B) + \mathcal{L}_{m}. 
\end{equation}
Where $\mathcal{L}_{m}$ represents matter Lagrangian. Through the addition of the boundary term $B$, $f(T)$ gravity becomes a generalized metric counterpart of $f(R)$ gravity \cite{no} since for $f(T,B) = f(-T+B) = f (R)$. \\
Varying the action \ref{1} with respect to the tetrad, I arrive at the following field equation, 
\begin{multline}\label{2}
16 \pi e \Theta^{\lambda}_{a}=  e B E ^{\lambda}_{a}f_{B}+2e E ^{\lambda}_{a} \square f_{B} - 2 e E ^{\sigma}_{a} \bigtriangledown^{\lambda}\bigtriangledown_{\sigma}f_{B}  + 4 \partial _{i} (e S_{a}^{i \lambda}) f_{T} \\ + 4 e \left[ \partial_{i}f_{B} + \partial_{i}f_{T}  \right] S_{a}^{i \lambda} - 4 e f_{T} T^{\sigma} _{ia}S_{\sigma}^{\lambda i} - e f E^{\lambda}_{a}. 
\end{multline}
In this work, I am interested in constraining some well motivated $f(T,B)$ gravity models in a flat FRW spacetime from the energy conditions.\\
To achieve this, I shall now assume a flat FLRW metric of the form 
\begin{equation}\label{3}
ds^{2} = dt^{2} - a(t)^{2}\sum_{i=1}^{3} dr^{2}_{i}
\end{equation} 
where $a(t)$ denote the scale factor. Since $f(T,B)$ gravity is not invariant under Lorentz transformations, care must be taken when tetrads are chosen. Undesirable cases like $f_{TT}=0$ appears when one considers a flat diagonal FLRW tetrad in spherical polar coordinates \cite{no}. \\
For the universe comprising predominantly of a perfect fluid, the Friedmann equations reads 
\begin{equation}\label{4}
\rho_{eff}(t) =\rho_{matter} -3 H^{2} (3 f_{B} + 2 f_{T}) - 3 \dot{H}f_{B} + 3 H \dot{f_{B}} + \frac{1}{2} f 
\end{equation}
and
\begin{equation}\label{5}
p_{eff}(t) =p_{matter} -\left[- (3 f_{B} + 2 f_{T}) (-3 H^{2} + \dot{H}) + \ddot{f_{B}} - 2 H \dot{f_{T}}  + \frac{1}{2} f\right],
\end{equation}
where $\rho_{matter}$ and $p_{matter}$ are the energy density and pressure for the matter fluid sources. The expressions of the energy density $\rho_{de}$ and pressure $p_{de}$ for the effective dark energy fluid in $f(T,B)$ gravity can be written as 
\begin{equation}
\rho_{de}(t) = -3 H^{2} (3 f_{B} + 2 f_{T}) - 3 \dot{H}f_{B} + 3 H \dot{f_{B}} + \frac{1}{2} f 
\end{equation}
and
\begin{equation}
p_{de}(t) = -\left[- (3 f_{B} + 2 f_{T}) (-3 H^{2} + \dot{H}) + \ddot{f_{B}} - 2 H \dot{f_{T}}  + \frac{1}{2} f\right].
\end{equation}

\section{Energy Conditions}\label{III}
Energy conditions are linear relationships based on Raychaudhuri equation comprising energy density and pressure employed to understand the nature of timelike, lightlike and spacelike curves and commonly used in studies related to singularities \cite{ft25,ft26}. The origin of these energy conditions is independent of any theory of gravity and are purely geometrical in nature \cite{fte,hawking}. The energy conditions are defined as \cite{fte16,fte17}
\begin{equation}\label{10}
NEC \Longleftrightarrow  \rho_{eff} (t) + p_{eff} (t)\geq 0.
\end{equation}
\begin{equation}
WEC\Longleftrightarrow  \rho_{eff} (t)\geq 0 \hspace{0.1in} \text{and} \hspace{0.1in}  \rho_{eff} (t) + p_{eff} (t)\geq 0. 
\end{equation}
\begin{equation}
SEC\Longleftrightarrow  \rho_{eff} (t) + 3 p_{eff} (t)\geq 0 \hspace{0.1in} \text{and} \hspace{0.1in} \rho_{eff} (t) + p_{eff} (t)\geq 0.
\end{equation}
\begin{equation}\label{11}
DEC \Longleftrightarrow  \rho_{eff} (t)\geq 0 \hspace{0.1in} \text{and} \hspace{0.1in} \rho_{eff} (t) \pm p_{eff} (t)\geq 0.
\end{equation}
In cosmology, the higher derivatives of Hubble parameter $H = \frac{\dot{a}}{a}$ such as deceleration, jerk and snap parameters can be defined as
\begin{equation}\label{6}
q = \frac{-1}{H^{2}}\frac{\ddot{a}}{a}, \hspace{0.15in} j = \frac{1}{H^{3}}\frac{\dddot{a}}{a}, \hspace{0.15in} s = \frac{1}{H^{4}}\frac{\ddddot{a}}{a},
\end{equation}
respectively.\\
Now, using Eq. \ref{6}, the derivatives of $H$ can be recast into following forms 
\begin{equation}\label{7}
\dot{H} = - H^{2} (1+q) 
\end{equation}
\begin{equation}\label{8}
\ddot{H} = H^{3}(2 + 3q + j ) 
\end{equation}
\begin{equation}\label{9}
\dddot {H} = H^{4} (-3 - 5 q - 2j + s ).
\end{equation}
The matter sector (i.e, $\rho_{matter}$ and $p_{matter}$) obey the energy conditions and that the validation or violation of the energy conditions is purely determined by the dark energy fluid (i.e, $\rho_{de}$ and $p_{de}$). Therefore using Eqs. \ref{7} - \ref{9} in Eqs. \ref{4} \& \ref{5}, the energy conditions (Eqs. \ref{10}-\ref{11}) can be re-expressed as 
\begin{multline}
NEC \Longleftrightarrow -H^2 (q+1) (3 f_{B}+2 f_{T})+3 f_{B} H^2 (q+1)-3 \dot{f_{B}} H^3 (q+1) \\ - \ddot{f_{B}} H^3 (j+3 q+2)-2 \dot{f_{T}} H^3 (q+1)+0 \geq 0.
\end{multline}
\begin{multline}
WEC \Longleftrightarrow 0.5 f-3 H^2 (-f_{B} (q-2)+ \dot{f_{B}} H (q+1)+2 f_{T}) \geq 0, \hspace{0.25in} \text{and} \\   -H^2 (q+1) (3 f_{B}+2 f_{T})+3 f_{B} H^2 (q+1)-3 \dot{f_{B}} H^3 (q+1)-\ddot{f_{B}} H^3 (j+3 q+2)-2 \dot{f_{T}} H^3 (q+1)+0 \geq 0.
\end{multline}
\begin{multline}
SEC\Longleftrightarrow 3 \left( \splitfrac{-0.5 f-H^2 (q+1) (3 f_{B}+2 f_{T})+3 H^2 (3 f_{B}+2 f_{T})}{-\ddot{f_{B}} H^3 (j+3 q+2)-2 \dot{f_{T}} H^3 (q+1)}\right)\\+0.5 f-3 H^2 (3 f_{B}+2 f_{T})+3 f_{B} H^2 (q+1)-3 \dot{f_{B}} H^3 (q+1)\geq 0,\hspace{0.25in} \text{and} \\ -H^2 (q+1) (3 f_{B}+2 f_{T})+3 f_{B} H^2 (q+1)-3 \dot{f_{B}} H^3 (q+1)-\ddot{f_{B}} H^3 (j+3 q+2)-2 \dot{f_{T}} H^3 (q+1)+0 \geq 0.
\end{multline}
\begin{multline}
DEC \Longleftrightarrow 0.5 f-3 H^2 (-f_{B} (q-2)+ \dot{f_{B}} H (q+1)+2 f_{T}) \geq 0, \hspace{0.25in} \text{and} \\ \left(0.5 f-3 H^2 (-f_{B} (q-2)+\dot{f_{B}} H (q+1)+2 f_{T})\right)\\ \pm \left(H^2 (-(3 f_{B} (q-2)+H (\ddot{f_{B}} (j+3 q+2)+2 \dot{f_{T}} (q+1))+2 f_{T} (q-2)))-0.5 f\right)\geq 0.
\end{multline}
\section{Constraints on $f(T,B)$ Gravity Models}\label{IV}
In this section I shall constrain some well motivated $f(T,B)$ gravity models from energy conditions. Note that current cosmological observations indicate $\omega = p_{de}/\rho_{de}= -1.03 \pm 0.03$ \cite{planck}. Hence, the SEC should violate under such conditions as $\rho_{de} + 3 \omega \rho_{de} < 0$ at the present epoch. Additionally, the WEC must assume values close to zero. Therefore, I focus primarily on the values of the free parameters in $f(T,B)$ gravity models for which the energy density remains positive and the WEC remains positive but very close to zero. For the analysis, I shall use $H_{0} = 0.692$, $q_{0}=-0.545$, $j_{0}=0.776$, and $s_{0}=-0.192$ \cite{cap}. 
\subsection{Power Law Model: $f(T,B)=\alpha  B^n+\beta  T^m$}
For the first case, I use the $f(T,B)$ gravity model of the form $f(T,B)=\alpha  B^n+\beta  T^m$ introduced in \cite{no} where $\alpha, \beta, n$ and $m$ are free parameters. For $n=1$, the model reduces to $f(T,B)=\alpha  B+\beta  T^m$ which is equivalent to a power-law $f(T)$ gravity model since $B$ is a boundary term and therefore any linear functional form of $B$ does not introduce changes in the field equations. Note that since there are four free parameters, constraints on the free parameters cannot be yielded analytically. Therefore, I assume some rational values for which the WEC remains positive but very close to zero. \\
The WEC for this model can be written as: 
\begin{multline}
WEC \Longleftrightarrow  0.5 \left(\beta  6^m \left(H^2\right)^m+\alpha  6^n \left(3 H^2-H^2 (q+1)\right)^n\right)\\ -3 H^2 \left(\beta  2^m 3^{m-1} m \left(H^2\right)^{m-1}+
 \alpha  2^{n-1} 3^n n \left(3 H^2-H^2 (q+1)\right)^{n-1}\right) \\ +\alpha  H^2 2^{n-1} 3^n n (q+1) \left(3 H^2-H^2 (q+1)\right)^{n-1}\\ +\alpha  H^3 \left(-2^{n-1}\right) 3^n (n-1) n (q+1) (6 H-2 H (q+1)) \left(3 H^2-H^2 (q+1)\right)^{n-2}\geq 0, 
 \hspace{0.1in}\\ \text{and} \\  -H^2 (q+1) \left(\beta  2^m 3^{m-1} m \left(H^2\right)^{m-1}+\alpha  2^{n-1} 3^n n \left(3 H^2-H^2 (q+1)\right)^{n-1}\right)\\ +\alpha  H^2 2^{n-1} 3^n n (q+1) \left(3 H^2-H^2 (q+1)\right)^{n-1}\\ +\beta  H^4 \left(-2^{m+1}\right) 3^{m-1} (m-1) m (q+1) \left(H^2\right)^{m-2}-  \alpha  H^3 6^{n-1} n (j+3 q+2)\\ \left(\splitfrac{ (n-2) (n-1) (6 H-2 H (q+1))^2 \left(3 H^2-H^2 (q+1)\right)^{n-3}}{+(n-1) (6-2 (q+1))  \left(3 H^2-H^2 (q+1)\right)^{n-2}}\right)-\\ \alpha  H^3 2^{n-1} 3^n (n-1) n (q+1) (6 H-2 H (q+1)) \left(3 H^2-H^2 (q+1)\right)^{n-2} \geq 0.
\end{multline}

For this model, upon substituting $\alpha=5$, $\beta=-0.1$, $n=0.1$ and $m=2.3$, the WEC $(\rho_{0}+p_{0})\simeq 0.09$ and therefore compatible with current observational constraints. 
\subsection{Mixed Power Law Model: $f(T,B)=\alpha  B^n T^m$}
For the second case, I use $f(T,B)=\alpha  B^n T^m$ introduced in \cite{no} where the authors adopted Noether Symmetry Approach to reconstruct the cosmological solutions. In this model $\alpha \neq 0$, $m \neq 0$ and $n$ are free parameters. The authors reported the constraint $n=1-m$ for which the cosmological solution is given as $a(t)=a_{0}t^{(1+n)/3}$. Trivial cases like $m=0$ should be avoided. For this model, I use the constraint on $n$ and $m$ reported in \cite{no} and check the viability of this cosmological model against energy conditions.  \\
The WEC for this model reads: 
\begin{multline}
WEC \Longleftrightarrow 0.5 \alpha  \left(H^2\right)^m 6^{m+n} \left(3 H^2-H^2 (q+1)\right)^n\\+\alpha  n (q+1) \left(H^2\right)^{m+1} 2^{m+n-1} 3^{m+n} \left(3 H^2-H^2 (q+1)\right)^{n-1}\\
-3 H^2 \left( \splitfrac{\alpha  m \left(H^2\right)^{m-1} 2^{m+n} 3^{m+n-1} \left(3 H^2-H^2 (q+1)\right)^n}{+ \alpha  n \left(H^2\right)^m 2^{m+n-1} 3^{m+n} \left(3 H^2-H^2 (q+1)\right)^{n-1}}\right)\\ -3 H^3 (q+1) \left(\splitfrac{ \alpha  H m n \left(H^2\right)^{m-1} 2^{m+n} 3^{m+n-1} \left(3 H^2-H^2 (q+1)\right)^{n-1}} {+ \alpha  (n-1) n  \left(H^2\right)^m (6 H-2 H (q+1)) 6^{m+n-1}   \left(3 H^2-H^2 (q+1)\right)^{n-2}}\right)\geq 0,   \hspace{0.3in}\\ \text{and} \\     \left(  \alpha  H \left(H^2\right)^m \left(-2^{m+n}\right) 3^{m+n-1}\right)     \left(  -H^2 (q-2)\right)^{n-1} \\ \times \left(\splitfrac{ n (j+3 q+2) \left(2 m^2+m (4 n-5)+2 n^2-5 n+3 \right)} { -H (q+1) \left( 2 m^2 (q-2)  +m (n (2 q-7)-q+2)-3 (n-1) n\right)}\right)\geq 0
\end{multline}
In this case, I substitute $\alpha=-1$, $m=0.2$ and $n=1-m=0.8$, to obtain WEC $(\rho_{0}+p_{0})\simeq 0.07$. The WEC in this case is obeyed and consistent with observations.  
\subsection{Logarithmic Model: $f(T,B)=\alpha  \log (B)+\beta  T$}
For the third case, I propose a novel cosmological model of the form $f(T,B)=\alpha  \log (B)+\beta  T$ where $\alpha \neq 0$ and $\beta \neq 0$ are free parameters. Since the model has only two free parameters, I find an analytical expression for which the WEC is obeyed. \\
The WEC for this model is given as:
\begin{multline}
WEC \Longleftrightarrow -3 H^2 \left(2 \beta +\frac{\alpha }{2 \left(3 H^2-H^2 (q+1)\right)}\right)+0.5 \left(6 \beta  H^2+\alpha  \log \left(6 \left(3 H^2-H^2 (q+1)\right)\right)\right)\\+\frac{\alpha  H^2 (q+1)}{2 \left(3 H^2-H^2 (q+1)\right)}+\frac{\alpha  H^3 (q+1) (6 H-2 H (q+1))}{2 \left(3 H^2-H^2 (q+1)\right)^2} \geq 0, \\ \text{and} \\
-H^2 (q+1) \left(2 \beta +\frac{\alpha }{2 \left(3 H^2-H^2 (q+1)\right)}\right)+\frac{\alpha  H^2 (q+1)}{2 \left(3 H^2-H^2 (q+1)\right)} \\ -\frac{1}{6} \alpha  H^3 \left(\frac{2 (6 H-2 H (q+1))^2}{\left(3 H^2-H^2 (q+1)\right)^3}-\frac{6-2 (q+1)}{\left(3 H^2-H^2 (q+1)\right)^2}\right) (j+3 q+2)\\+\frac{\alpha  H^3 (q+1) (6 H-2 H (q+1))}{2 \left(3 H^2-H^2 (q+1)\right)^2} \geq 0 .
\end{multline}
Upon substituting the respective values of $H_{0}$, $j_{0}$ and $s_{0}$, the constraints on $\alpha$ and $\beta$ for the validation of the WEC ($\rho_{0}+p_{0} \geq 0$) reads
\begin{equation}
\frac{\alpha}{\beta}\gtrsim -0.2
\end{equation}
For this model, I substitute $\alpha=0.1$ and $\beta =-0.5$ to obtain $\rho_{0}+p_{0}\simeq 0.0017$ which is in excellent agreement with observations. 
\section{Conclusions}\label{V}
Recent cosmological observations make it clear-cut that the universe is undergoing an accelerated expansion by virtue of the so-called dark energy. Furthermore, galactic rotational curves and gravitational lensing experiments hint at the existence of a massive abundance of non-baryonic dark matter. There is no general consensus about the nature of these enigmatic quantities which together makes about 95\% of the energy budget of the universe. To explain these eccentric cosmological observations, viable cosmological models which are either adhered to general relativity or infra-red modifications of general relativity have been proposed and are currently being investigated to understand their viability and applicability. \\
Modified theories of gravity are geometrical extensions of general relativity and have widespread use in modern cosmology. The cosmological models reconstructed upon these extended theories of gravity have been able to explain inflation, baryon asymmetry, coincidence problem and the cosmic acceleration without the assistance of dark energy. Interestingly, these models under certain conditions can also avoid the initial big bang singularity and explain the flat rotational curves without requiring dark matter. Many of these extended theories of gravity have been constrained from the big-bang nucleosynthesis, redshift drift, supernovae, Planck and other datasets to obtain the range in parameter spaces. \\
In this work, I investigate the viability of some well motivated $f(T,B)$ teleparallel gravity models from the energy conditions. $f(T,B)$ is a recently proposed straightforward generalization of the popular $f(T)$ teleparallel gravity theory by the incorporation of a boundary term $B=\frac{2}{e}\partial_{i}(e T ^{i}) = \bigtriangledown_{i}T^{i}$.  Through the addition of the boundary term $B$, $f(T)$ gravity becomes a generalized metric counterpart of $f(R)$ gravity since for $f(T,B) = f(-T+B) = f (R)$ \cite{no}.\\
I find that for all the cosmological models investigated in this study, there exists corners in parameter spaces for which the weak energy condition attains a minute positive value to suffice the late-time acceleration. Note that every model under consideration predicts a violation of the strong energy condition at the present epoch. \\
However, as first reported in \cite{ecfr} and further discusssed in \cite{ecfg}, issues regarding the application of energy conditions in modified theories of gravity is an open interrogation, which eventually relates to the confrontation between theory and observations.  \\
As a possible extension, readers are encouraged to employ the constraints on model parameters obtained from the work in other cosmological scenarios such as gravitational baryogenesis, inflation and late time acceleration to investigate the cosmological viability and applicability of $f(T,B)$ gravity in the era of precision cosmology.
\section*{Acknowledgments}
I thank the referee for his or her suggestions.

\end{document}